\documentstyle[prb,aps]{revtex}
\begin{document}
\title{Surfactants in semiconductor heteroepitaxy: Thermodynamics 
and/or kinetics?}
\author{Ivan Markov$^*$}
\address{Institute of Physical Chemistry, Bulgarian Academy of Sciences, 
1113 Sofia, Bulgaria}
\maketitle

\begin{abstract}
The effect of surfactants on the thermodynamics and kinetics of semiconductor 
heteroepitaxy is briefly discussed. It is argued that the way the surfactants 
suppress the thermodynamic driving force for 3D islanding depends on the 
mechanism of exchange of overlayer and surfactant atoms. If the overlayer 
atoms occupy bulk-like positions provided by the outgoing surfactant dimers, 
as is the case of Ge/Si(001), large atomic displacements are forcibly 
inhibited, and the wetting of the overlayer by the substrate becomes nearly 
complete. This complete wetting of the overlayer by the substrate means 
a zero thermodynamic driving force for 3D islanding. Once the thermodynamics 
do not require 3D islanding a change of the growth mode with the temperature 
should not be observed. Thus, the temperature change of the growth mode 
appears as an indication for the primary role of the kinetics, as in the 
case of Ge/Si(111).

\end{abstract}

\twocolumn
\section{INTRODUCTION}

The presence of surface active species (surfactants) on the surface of
growing crystals changes dramatically the mode of growth, the rate of
nucleation, the transition from step flow to two-dimensional (2D)
nucleation, the structure of steps, the distribution of stress, surface
alloying, the concentration of defects, etc.\cite{Zinke} The major effect
is the suppression of the three-dimensional (3D) islanding so that smooth
films suitable for fabrication of microelectronic devices could be produced.
Although the initial attempts to understand the effect of the surfactants on
the mode of growth were based on thermodynamic grounds,\cite{Copel} there is
still a debate in the literature between the adherents of the thermodynamic
and the kinetic views. That is why we will make an attempt to consider 
both the thermodynamic and the kinetic aspects of the effect of
surfactants on the thin film growth mode. We will do that in the most
general way without accounting for the particular properties of the
materials. We note only that in the case of semiconductor growth a complete 
monolayer of the surfactant is needed.\cite{Daneft}

\section{THERMODYNAMICS}

If we want to change the direction of a process we have to change the sign
of the thermodynamic force that drives it. If we want to prevent a
particular process to take place we have to suppress the corresponding
thermodynamic driving force making it as close to zero as possible. The
thermodynamic driving force which determines the occurrence of one or
another mechanism of epitaxial growth is the difference $\Delta \mu =\mu
(n)-\mu _{3D}^{0}$ of the chemical potential, $\mu (n)$, of the overlayer
which depends on the film thickness measured in number $n$ of monolayers 
counted from the interface, and the chemical potential, $\mu 
_{3D}^{0}$, of the bulk 3D crystal.\cite{Stoyan,Mark5} The thickness 
dependence of the film chemical potential $\mu (n)$ originates mostly from 
the thickness distribution of the strain due to the lattice misfit, but the 
interaction between the deposit and the substrate, which rapidly decreases 
with the distance from the interface and could be neglected beyond several 
monolayers should be also accounted for.\cite{Stoyan,Mark5} If we deposit a 
crystal $A$ on the surface of a crystal $B$ in absence of a surfactant the 
thermodynamic driving force can be written in terms of surface 
energies $\sigma _A$ and $\sigma _B$\cite{Mark1} 
\begin{equation}\label{3sigma}
\label{bauer}\Delta \mu =a^{2}[{\sigma _{A}+\sigma _{AB}(n)-\sigma _{B}}] 
\end{equation}
where $a^{2}$ is the area occupied by an atom, and $\sigma _{AB}(n)$ is the 
interfacial energy which includes in itself the misfit strain energy and the 
attenuation of the energetic influence of the substrate.\cite{Mark1} In fact, 
this is the familiar 3-$\sigma $ criterion of Bauer.\cite{Bauer} 

We can write the above expression in terms of interatomic 
energies\cite{Kaish}
\begin{equation}\label{a-b}
\mu (n)=\mu _{3D}^{0}+(E_{AA}-E_{AB}) = \mu _{3D}^{0}+E_{AA}\it \Phi
\end{equation}
where $E_{AA}$ and $E_{AB}$ are the energies per atom to disjoin a 
half-crystal $A$ 
from a like half-crystal $A$ and an unlike half-crystal $B$. It is in fact 
the adhesion energy $E_{AB}$ which includes in itself the thickness 
distribution of the strain energy. In the above equation ${\it \Phi } = 
1-E_{AB}/E_{AA}$ is the so called adhesion parameter which accounts for the 
wetting of the overgrowth by the substrate. This is the same parameter which 
enters the work of formation of nuclei on a foreign substrate in the 
classical nucleation theory.

Equations (\ref{3sigma}) and (\ref{a-b}) are identical. Eq. (\ref{3sigma}) 
can be readily obtained from (\ref{a-b}) by using the definition of surface 
free energy and the relation of Dupr\'e.\cite{Mark1} The essential physics, 
however, is in Eq. (\ref{a-b}) as the surface energies are derivatives of the 
interatomic forces. It shows that {\it the thermodynamic driving 
force for occurrence of one or another mode of growth is the relative 
adhesion/cohesion difference, or in other words, the wetting}. The 
incomplete wetting, $0 < {\it \Phi } < 1$, is the driving force for 3D 
nucleation and growth, whereas the complete wetting, ${\it \Phi } \le 0$, is 
the driving force for planar growth by nucleation and growth of 2D monolayer 
height nuclei.\cite{Kaish} 

When $0 < {\it \Phi } < 1$ $\mu (n)>\mu _{3D}^{0}$, and 3D islands are formed 
on top of the substrate surface from the very beginning of deposition. This 
is the well known Volmer-Weber (VW) growth and $\mu (n)$ goes asymptotically 
to $\mu _{3D}^{0}$ from above (the curve denoted by VW in Fig. 
\ref{chempot}).\cite{Stoyan,Mark5,Mark1}

In the other extreme of complete wetting (${\it \Phi } \le 0$) either 
layer-by-layer (Frank-van der Merwe 
or FM) growth at negligible lattice misfit, or a Stranski-Krastanov (SK) 
growth (layer-by-layer growth followed by 3D islands) at a perceptible 
misfit, are thermodynamically favored.\cite{Kern0,Gilmer} In the case of FM 
growth $\mu (n)$ goes asymptotically to $\mu _{3D}^{0}$ from below (the 
curve denoted by FM in Fig. \ref{chempot}). The first monolayer has the 
lowest chemical potential owing to the strongest interaction with the 
substrate, and a second monolayer can form only after the completion of the 
first one.

The SK growth mode appears as a result of the interplay of the film -
substrate bonding, strain, and surface energies. A wetting layer consisting
of an integer number of monolayers is first formed by a FM
mode which is driven by a negative thermodynamic driving force $\Delta \mu $
(complete wetting). Further FM growth becomes unfavorable as the strain 
energy accumulates linearly with film thickness. In addition, the stronger 
attraction from the substrate, that overcompensates the strain energy, 
disappears beyond several atomic diameters. On
top of the wetting layer 3D islands form and grow under the influence of
a positive thermodynamic driving force (incomplete wetting). If misfit
dislocations (MDs) are introduced immediately after the completion of the 
wetting layer unstrained 3D islands are formed and grow on top of it. This 
is the classical Stranski-Krastanov growth in which the work needed to create 
new surfaces is overcompensated by a complete strain relaxation. It was found
recently that initially coherent (dislocation-free) 3D islands are formed
(coherent Stranski-Krastanov growth),\cite{Eagl,Leo,Vinh} in which the
formation of a new surface is overcompensated by a gradual strain
relaxation. In both cases the 3D islands and the wetting layer represent
necessarily different phases separated by an interphase boundary. This
boundary is determined by the displacements of the atoms belonging to the 
first atomic plane of the islands from
the bottoms of the corresponding potential troughs of the uniformly strained
wetting layer. In the classical SK growth the displacements are described  
in terms of MDs,\cite{Mat1} whereas in the coherent SK 
growth the atoms that are closer to the islands edges are displaced owing to 
the islands finite size.\cite{Kormar} These displacements give rise to 
weaker adhesion, or in other words, to incomplete wetting. It is namely this 
incomplete wetting (or weaker adhesion) which appears as the thermodynamic 
driving force for the 
3D islanding in the SK mode of growth. In other words, we can treat the SK 
mode as a FM mode driven by complete wetting ($\Delta \mu < 0$), followed by 
VW mode driven by incomplete wetting ($\Delta \mu > 0$) (the curve denoted 
by SK in Fig. \ref{chempot}). 

When considering the effect of the surfactant on the mode of growth we 
note that the chemical potential $\mu _{3D}^{0}$ of the infinitely large 
crystal does not depend on whether there are impurities adsorbed on its 
surface or not.\cite{Bliz,Pang} This, however, is not true for sufficiently 
thin films. If a monolayer thick film of $A$ is covered by a monolayer of a 
surfactant (S) atoms the growth mode criterion (\ref{bauer}) turns 
into\cite{Paun} 
\begin{equation}\label{gibbs}
\Delta \mu _{S}=a^{2}(\sigma _{AB}+\sigma _{SA}-\sigma _{SB}) 
\end{equation}
where the $\sigma $'s are the corresponding interfacial energies. This 
expression is equivalent to
\begin{equation}\label{paun}
\mu _{S}(n)=\mu _{3D}^{0}+(E_{AA}-E_{AB})-(E_{SA}-E_{SB}) 
\end{equation}
where $\mu _{S}(n)$ is the chemical potential of the film covered by S atoms. 
It is immediately seen that if 3D islanding is thermodynamically favored in 
absence of a surfactant, i.e. $E_{AA}>E_{AB}$, the inequality (\ref{paun}) 
may have the opposite sign if the adhesion of the surfactant is stronger to 
the overlayer rather than to the substrate and thus the third term in 
(\ref{paun}) overcompensates the second one. Then the surfactant is ''good'' 
from thermodynamic point of view if it adheres more strongly to the overlayer, 
thus changing the sign of the thermodynamic driving force.

This is valid, however, only for the transition from VW to FM growth as is
the case of the As mediated growth of Si on Ge(001).\cite{Copel,Tak} Even in 
this case it is valid only at zero misfit. After several monolayers of $A$ 
the energetic contact between the surfactant and the substrate $B$ is 
completely lost ($E_{SB} \rightarrow E_{SA}$), and if the misfit 
is large 3D islands should appear resulting in SK mode. In systems 
that follow the SK growth mode where, beyond the wetting layer, $A$ is 
deposited on strained $A$,\cite{Orr} the difference $E_{SA}-E_{SB}$ will be 
nearly equal to zero. Many papers both 
theoretical\cite{Tromp,Yu,Jiang,Jenk} and 
experimental\cite{Cao,Yu1,Bosh1,Thorn,Ost1,Ide} have been devoted 
to the problem considering mainly the surfactant growth of Ge on Si(001). 
However, in most of the papers quoted above the mechanism of the exchange of 
the surfactant (As or Sb) atoms with Ge atoms belonging to the first 
monolayer on the Si(001) surface have been studied.
It is obvious, that such studies cannot explain the effect of the surfactant
to suppress the 3D islanding as the first monolayer belongs to the wetting
layer and it will grow in a FM mode pseudomorphous with the substrate
irrespective of there is a surfactant on top or not.

Obviously, in the SK mode the surfactant should make the thermodynamic 
driving force for 3D islanding on top of the wetting layer equal to zero for 
considerable film thickness. As mentioned above in the SK mode the atoms 
which are in contact with the wetting layer should be displaced from their 
''bulk-like'' positions either to make MDs or near to the islands edges. 
These displacements make the wetting incomplete, or which is the same, make  
the adhesion weaker than the cohesion ($E_{AB}<E_{AA}$), which is the 
thermodynamic driving force for 3D islanding. As found in a series of papers 
the surfactant group-V atoms form dimers on Si(Ge)(001) which reside exactly 
where the Si(Ge) dimers should be located.\cite{Cao,Bosh1,Nogami,Uhr,Beck} 
After the surfactant and overlayer dimers exchange places the latter occupy 
the epitaxial sites that are provided by the outgoing surfactant. The 
surfactant dimers on top do not permit large displacements of the overlayer 
atoms underneath, and in turn incomplete wetting. In other words, the 
surfactant atoms press the overlayer atoms into bulk-like positions and do 
not allow the formation of misfit dislocations. Thus, {\it the surfactants 
suppress the thermodynamic driving force for 3D islanding by not allowing 
large atomic displacements}. Once the thermodynamics do not require 3D 
islanding a change of the growth mode with the temperature should not be 
observed.

\section{KINETICS}

In absence of a surfactant the atoms arriving at the crystal surface diffuse 
on it, join pre-existing steps or islands, or give rise to new 
islands. At sufficiently 
high temperatures the diffusivity of the atoms on the terraces is high and 
they reach the pre-existing steps before meeting with each other. This 
results in a step-flow growth. At low temperatures and in turn low 
diffusivity the atoms meet with each other before reaching the steps and 
give rise to 2D nuclei on the terraces. In presence of a complete monolayer 
of S atoms two new phenomena take place. First, the surfactant changes the 
energetics of incorporation of atoms into ascending and descending steps, 
and second, the overlayer atoms have to exchange places with the S atoms on 
the terraces or at the steps in order to join the crystal lattice. Both 
phenomena lead to a change of the diffusivity of the adatoms across the 
steps and on terraces, and in turn exert an effect on the direction of 
transport (upward or downward) of atoms. Moreover, the change of diffusivity 
on terraces affects strongly the kinetics of nucleation and growth of the 
islands thus stimulating either the step flow or 2D nucleation. We consider 
in more detail the effect of the surfactant on the kinetics of attachment 
and detachment of atoms to and from the steps, and the exchange-deexchange 
kinetics on terraces.

Whereas an atom approaching an ascending step joins it upon striking, an
atom approaching a descending step has to overcome an additional
Ehrlich-Schwoebel (ES) barrier.\cite{Ehrl,Sch} If the ES barrier is low the 
overlayer grows more or less in a layer-by-layer mode. Otherwise, 
instabilities of different kind appear.\cite{Politi} Zhang and Lagally 
pointed out the possibility of a reduction of the ES 
barrier when overlayer atoms exchange sites with S atoms that decorate the 
steps, in such a way that the last barrier before descending is shifted 
downwards below the level of the surface diffusion barrier.\cite{ZL} The 
latter is in good agreement with first principle molecular dynamics 
calculations of displacement of Sb dimer at the step edge by a Si dimer 
through a push-out mechanism.\cite{Oh} In addition, Markov has suggested that 
atoms that approach ascending steps have to overcome an extra energetic 
barrier owing to the necessity to displace the S atoms already adsorbed at 
the step edges.\cite{Mark3} These effects which are of purely kinetic origin 
can suppress the thermodynamic driving force for 3D islanding in 
heteroepitaxial growth.

When thermodynamics predict 3D islanding, ($\Delta \mu > 0$) the atoms are 
more strongly bound to the upper layers rather than to the
lower layers. In other words the upper layers have lower chemical potentials 
(Fig. \ref{chempot}).\cite{Stoyan,Mark5} This results in a gradient of 
the chemical potential that drives the atoms upwards and is the driving 
force for the 2D-3D transformation.\cite{Stoyan,Mark5,Mark1} The appearance 
of an additional barrier at the ascending steps, and the reduction 
of the ES barrier at the descending steps owing to the decoration of the 
steps by S atoms, reverse the asymmetry of the attachment - 
detachment kinetics and thus lead to a suppression of the thermodynamic 
driving force for 2D-3D transition. A reverse gradient of adatoms of a {\it 
kinetic origin} appears that drives the atoms downwards. The latter results 
in a planar rather than a 3D growth.\cite{Mark3,Mark4} Thus the attachment - 
detachment kinetics suppress the thermodynamic driving force for 3D 
islanding. At high temperature the thermodynamics prevail and SK 
growth takes place irrespective of the presence of the surfactant. 
Surfactant induced FM growth occurs at lower temperatures. Note that this 
kinetic effect does not require a complete monolayer of the surfactant 
but a small amount which is sufficient to decorate the steps. 

In the presence of a complete monolayer of the surfactant a new kinetic
effect appears owing to the necessity of the overlayer atoms to join the
crystal lattice and of the S atoms to float over the surface. These
are the phenomena of exchange and deexchange of overlayer and S atoms. 
Zhang and Lagally\cite{ZL} first assumed that an exchange process between 
overlayer and S atoms on the terraces should take place on the S precovered
surfaces. Kandel and Kaxiras assumed later that a deexchange process is also 
possible and can play a significant role in 
epitaxial growth.\cite{Daneft,Dan1} They calculated the values of 0.8 and 
1.6 eV for the activation energies, $E_{ex}$ and $E_{dex}$,  for exchange 
and deexchange processes, respectively. Ko, Chang and Yi computed
through first-principles pseudopotential total-energy calculations the
values of the barriers for surface diffusion, exchange and deexchange for Ge
on Ga, As and Sb precovered Si(111) surface. They found that in all cases 
$E_{dex} > E_{ex}$ but the differences are not as large as the ones 
calculated by Kandel and Kaxiras.\cite{Ko} 

We introduce two time constants which characterize the exchange and
the deexchange processes both being normalized to the time, $t_1=1/F$,
necessary for deposition of a complete monolayer, $F$ being the the atom 
arrival frequency.\cite{Mark7,Mark8} First,
this is the mean residence time of the atoms on top of the S layer
before exchange $\tau _{ex}=(F/\nu )\exp (E_{ex}/kT)$ where $\nu $ is the 
attempt frequency. Second, this is the mean residence time before deexchange 
of an atom embedded in (or under) the S layer $\tau _{dex}=(F/\nu 
)\exp (E_{dex}/kT)$. Obviously, if a particular time constant is greater than 
unity the corresponding process will not occur. As 
$\tau _{dex} \gg \tau _{ex}$ three possibilities exist. The 
first is $\tau _{dex} \ll 1$. This extreme describes the {\it reversible 
exchange} as the adatoms have sufficient time to go back on top of the 
S layer through a deexchange process. A dynamic {\it exchange - 
deexchange equilibrium} is established. With the value of $E_{dex} = 1.6$ eV 
calculated by Kandel and Kaxiras we find, at 600K, $\tau _{dex} \approx 
0.002$. Then the exchange - deexchange equilibrium is established in the very 
beginning of the deposition process.\cite{Mark7,Mark8} The nucleation process 
takes place in a more or less disordered (because of the different size of 
the atoms involved) 2D phase - a monolayer consisting of mixed S and 
overlayer atoms. A considerable fraction of the incoming atoms 
remains on top of the S layer and diffuse fast to the pre-existing steps. The 
formation, growth and decay of nuclei take place through exchange and 
deexchange processes. The second extreme is $\tau _{ex} \ll 1 < \tau _{dex}$. 
This is the case of 
{\it irreversible exchange}. The incoming atoms rapidly exchange places with
S atoms and remain buried under the S layer. All processes 
of diffusion, nucleation and incorporation into islands and steps occur under 
(or between) the S atoms. The case $1 < \tau _{ex} \ll \tau _{dex}$ 
should be excluded as it means that the S atoms will remain buried 
under the arriving overlayer atoms.

We calculate further the nucleus density, $N_S$, in both cases of reversible 
and irreversible exchange. This is extremely important because of two 
reasons. First, the large number of small islands will promote layer-by-layer 
growth owing to the existence of a critical island size for second layer 
nucleation.\cite{StMar,Tersoff,Krug,Stefan} Second, higher nucleation rate 
leads to growth by 2D nuclei, whereas in the opposite case step-flow growth 
will take place. In all cases the expression for $N_S$ can be written in the 
form 
\begin{equation}\label{ns}
N_S=N_{0}\exp \bigl(-\frac \chi iE_S/kT\bigr)
\end{equation}
where $E_S$ combines all energy contributions which depend on the presence 
of the surfactant,\cite{Mark7,Mark8} and
\begin{equation}\label{n0}
N_{0} = \bigl(\frac{\nu }{F}\bigr)^{-\chi }\exp\bigl(\frac{\chi 
}{i}E_{0}/kT\bigr)
\end{equation}
is the nucleus density in absence of a surfactant ($E_S = 
0$),\cite{Mark7,Mark8} Thus, the island density scales with the ratio $\nu 
/F$, the scaling exponent $\chi $ being a function of the number $i$ of 
atoms in the critical nucleus.\cite{Ven,StDim} In the case of low barriers 
for attachment of atoms to 2D nuclei, the latter ''feel'' each other through 
the diffusion fields around them, and\cite{Ven,StDim} 
\begin{equation}\label{chiold}
\chi =\frac {i}{i+2}. 
\end{equation}

Kandel has recently shown by using a rate equation approach that in presence 
of surfactants which give rise to barriers for incorporation of atoms to the 
critical nuclei\cite{Dan2} 
\begin{equation}\label{chinew}
\chi = \frac{2i}{i+3}. 
\end{equation}
The scaling exponent (\ref{chinew}) has been independently derived by Markov 
by using a different approach.\cite{Mark6} He showed that it is
characteristic for kinetic regime of growth where the adatom concentration
between the islands is practically constant and the islands "do not feel"
the presence of each other. 

As has been noted in Ref. (\cite{Dan2}) the scaling exponent (\ref{chiold}) 
varies with $i$ from 1/3 to 1, whereas (\ref{chinew}) has values larger than 
unity already at $i>2$. Thus, one could distinguish between diffusion and 
kinetic regimes of growth if $\chi $ is smaller or greater than unity. 
Typical examples of the scaling exponent (\ref{chinew})
are the homoepitaxy of Si on Sn precovered surface of Si(111),\cite{Takaya}
and of Ge on Pb precovered surface of Si(111).\cite{TTT}

It is immediately seen that the exponential multiplying $N_0$ in Eq. 
(\ref{ns}) can be smaller or greater than unity depending on the interplay of 
the energies involved in $E_S$. If $E_S < 0$ the surfactant will 
stimulate the nucleus formation ($N_{S} \gg N_0$). The latter means that a 
greater density of smaller islands will be formed, and the film will grow by 
2D nucleation rather than by step-flow. What is more important is that the 
greater density of smaller nuclei leads to layer-by-layer growth rather than 
to the mound formation. In the opposite case the surfactant will inhibit the 
nucleation and will promote the step-flow growth. In the case of reversible 
exchange $E_S$ contains the term $-i[(E_{dex}-E_{ex}) - \Delta E_{sd}]$ 
where $\Delta E_{sd}=E_{sd}^{\circ}-E_{sd}$ is the the difference between 
the barriers for diffusion on the clean substrate, $E_{sd}^{\circ}$, and on 
the surfactant passivated surface, $E_{sd}$, ($E_{sd}^{\circ}>E_{sd}$). If 
$\Delta E_{sd}$ overcompensates the difference $E_{dex}-E_{ex}$, we could 
expect that $E_S > 0$, the nucleus formation will be inhibited, ($N_{S} \ll 
N_{0}$), and surfactant-induced step-flow growth will take 
place.\cite{Mark7,Mark8}

In the case of irreversible exchange Eq. (\ref{ns}) is still valid but the 
barriers $E_{dex}$ and $E_{ex}$ do not enter the energy term $E_S$. The
largest term in $E_S$ is the increment of the surface diffusion barrier 
$-i(E_{sd}-E_{sd}^{\circ})$ where now $E_{sd}>E_{sd}^{\circ}$ owing to the 
fact that the process of diffusion under the S layer is inhibited compared 
with that on the clean surface. It is unlikely this term to be 
overcompensated by the smaller positive terms, and hence, we can always 
expect that the surfactant will stimulate the 2D nucleation rather than the 
step-flow growth. 

Finally, we discuss briefly the reasons that lead to a change of the island 
density. Usually the latter is attributed to the change of atoms  
diffusivity. However, the change the island density might be due to other 
reasons. The nucleation rate depends on the Gibbs free energy for nucleus 
formation which is in turn a function of the nucleus size. The presence of 
the surfactant could lead to a change of the number of atoms in the critical 
nucleus.\cite{Mark2} As a result the values of $E_S$ and $E_{0}$ in Eqs. 
(\ref{ns}) and (\ref{n0}), as well as the scaling exponent $\chi $ will 
change in one and the same temperature interval. In addition, the regime of 
growth can change from diffusion to kinetic, so that the scaling exponent 
(\ref{chiold}) will be replaced by (\ref{chinew}). In this case, the island 
density will increase by orders of magnitude even if $i$ remains constant.

\section{GROWTH OF Ge ON Si}

We discuss in more detail a "model" system such as Ge/Si. As pointed out 
by Kaxiras,\cite{Kax} one and the same surfactant can have qualitatively 
different effect depending on the crystallographic orientation of 
the substrate. On Si or Ge(001) the surfactants form dimers which do not 
break during the exchange 
process.\cite{Tromp,Yu,Jiang,Jenk,Cao,Yu1,Bosh1} On (111) surface 
three different geometries (substitutional, trimers, and zig-zag chains) of 
the group-V adsorbates (P, As, Sb) are possible.\cite{Kax} In the first 
$1 \times 1$ structure one monolayer of S atoms is incorporated 
into the upper part of the (111) bilayer, whereas in the last two structures 
a monolayer of S atoms is bonded on top  of a complete (111) 
bilayer. In addition, the growth of the two surfaces of Si or Ge 
differs drastically. Whereas the (001) surface grows by single 
monolayers,\cite{Grif} the (111) surfaces grow by bilayers.\cite{Udi} 
Thus, we have completely different mechanisms of exchange of S and 
overlayer atoms. In the case of (001) surfaces the growth unit is a dimer so 
that an exchange of S and overlayer dimers takes place through a 
single monolayer.\cite{Tromp,Yu,Jiang,Jenk,Cao,Yu1,Bosh1} On (111) 
surfaces the mechanism of exchange depends on the surface geometry of the 
surfactant. For example, if the surfactant geometry is substitutional we 
could suppose that single overlayer atoms should displace single S atoms from 
their positions in the upper monolayer that constitutes the bilayer. After 
the complete building of the upper monolayer atoms belonging to the next 
lower monolayer should be incorporated first in order to serve as a template 
for the S atoms in the next upper monolayer. Thus one surfactant monolayer 
should be displaced by a complete bilayer of the overgrowth the process 
taking place by exchange of single atoms. Comparing both surfaces we could 
speculate that on (111) surface the surfactant could not suppress 
successfully the thermodynamic driving force for 3D islanding by not allowing 
large atomic displacements (formation of misfit dislocations). 

We consider first the growth of Ge on Si(111). Hwang, Chang and Tsong studied 
this system by using Pb as a surfactant. They established that the Pb atoms 
make a substitutional $1 \times 1$ structure on top of the Ge surface. It was 
found that the saturation nucleus density in the submonolayer regime scales 
with the deposition flux with $\chi  = 1.76$.\cite{TTT} 
This clearly shows that in the Pb mediated epitaxy of Ge on Si(111) the 
scaling exponent is given by Eq. (\ref{chinew}) and the growth proceeds in 
a kinetic regime. In other words, significant energy barriers appear for 
attachment of atoms to the critical nuclei owing to the necessity of 
exchange of overlayer and S atoms. The same scaling exponent ($\chi  = 1.8$) 
has been obtained in the homoepitaxial growth of Si on Sn precovered surface 
of Si(111),\cite{Takaya} after comparing the experimental data for the 
critical terrace width for step flow growth with the corresponding 
theory.\cite{Mark7,Mark8} Under clean conditions in the same system 
Voigtl\"ander et al. obtained a scaling exponent smaller than unity 
($\chi  = 0.85$).\cite{Bert}

The observations of Hwang et al. show unambiguously that Pb has a prominent 
kinetic effect on the submonolayer growth of Ge on Si(111).\cite{TTT} In a 
further study the same authors found that 3D islanding is suppressed below 
470K. At temperatures higher than 470K 3D islanding takes 
place.\cite{ttt1} This observation could be explained by the suppression of 
the thermodynamic driving force for 3D islanding by the attachment - 
detachment kinetics mentioned above. The upward surface flux is hindered 
whereas the downward flux is enhanced. We could conclude that the 
suppression of 3D islanding in Ge growth on Pb precovered Si(111) is due 
more to kinetic rather than to thermodynamic reasons. 

Voigtl\"ander and Zinner studied the growth of Ge on Si(111) mediated by 
Sb.\cite{Voigt} They established the same substitutional structure of the Sb 
atoms on the surface of Ge. They found that a transition from surfactant 
mediated layer-by-layer growth to the equilibrium SK mode takes place 
above twice higher temperatures (900K) as compared with the Pb mediated 
growth. The higher temperatures could be related to stronger chemical 
bonding but qualitatively the surfactant effect is the same.

The system Ge/Si(001) is the most studied one.\cite{Daneft} Ide has observed 
that after annealing at about 800K for 10 min the 2D Ge islands formed in 
submonolayer regime on As precovered Si(001) did not change whereas the 
islands deposited under clean conditions disappeared.\cite{Ide} This 
observation shows that 2D islands formed under the surfactant layer are 
more stable than those without S atoms on top of them. 
Antimony mediated deposition and annealing experiments carried out by Osten 
et al. showed that Sb prevents 3D islanding, but also can smooth an already 
islanded film.\cite{Ost1} Katayama et al.\cite{Katay} found that as the 
coverage of the surfactant increases, the intermixing of Ge and Si, as well 
as the nucleation and the growth of macroscopic Ge islands, are suppressed. 
Both observations could be explained with the forced occupation of 
epitaxial sites by the Ge atoms under the influence of the S dimers on top 
of them. On the other hand, Ide also established 
that at 800K the Ge atoms reach and join the preexisting steps of Si (step 
flow growth) on the clean surface, whereas 2D islands are formed in the 
presence of As.\cite{Ide} The latter unambiguously demonstrates the kinetic 
effect of the surfactant on the mode of growth by decreasing the diffusivity 
of the incoming atoms. 

Most of the studies of surfactant mediated growth of Ge on Si(001) surface 
are carried out at or near to 
500$^{\circ}$C.\cite{Copel,Cao,Yu1,Thorn,Ide,Copel1,Franc} To the author's 
knowledge higher temperature deposition has been carried out by Tromp and 
Reuter (630$^{\circ}$C),\cite{Tromp} and by Horn von Hoegen et al.\cite{Horn} 
at 700$^{\circ}$C. No 3D islanding except for mound formation has been 
noticed on top of the wetting layer at the highest temperature 
used.\cite{Horn} A change of the growth mode with temperature as in the case 
of Ge/Si(111) has not been reported so far.\cite{Horn2} This can be 
interpreted as follows. The 
surfactant suppresses successfully the thermodynamic driving force for 3D 
islanding by not allowing large atomic displacements and the growth 
continues in a layer-by-layer mode (with the unavoidable roughness to 
relieve the strain) untill MDs are introduced at the 
interface. Once the thermodynamics do not require 3D islanding a change of 
the growth mode with the temperature should not be observed.
 
Jenkins and Srivastava carried out first principle density functional theory  
(DFT) calculations of the 
mode of growth of Ge on Sb precovered Si(001).\cite{Jenk} They 
studied the behavior of the chemical potential of 
the Ge film relative to the bulk chemical potential, $\mu _{3D}^{0}$, in 
absence and presence of Sb as a surfactant. They found that in absence of Sb 
the chemical potentials of the first three monolayers of Ge, which belong to 
the wetting layer, are smaller (more negative) than $\mu _{3D}^{0}$ (complete 
wetting). The chemical potential of the forth monolayer becomes more positive 
than $\mu _{3D}^{0}$ (incomplete wetting) which is an indication of 3D 
islanding after the first three monolayers. In presence of Sb the chemical 
potential of the first monolayer is slightly more negative that 
$\mu _{3D}^{0}$ but the chemical potentials of the remaining three monolayers 
are equal to $\mu _{3D}^{0}$ within the accuracy 
of the calculations. An error is introduced in the calculations by not 
allowing the Ge overlayers to relax laterally. Thus after the exhange of 
Ge and Sb dimers the Ge atoms occupy the bulk-like positions left by the 
outgoing Sb dimers and the surfactant successfully suppressed the 
thermodynamic driving force for 3D islanding. In fact, the surfactant 
replaced the curve SK by the curve FM in Fig. (\ref{chempot}).

Similar first principle DFT calculations have been carried out by 
Gonz\'a- les-M\'endez and Takeuchi for the growth of Si on Ge(001) by using As 
as a surfactant.\cite{Tak} In absence of a surfactant Si grows in VW mode as 
3D islands directly on the Ge surface.\cite{Maree} In presence of As the 
overlayer prefers to grow in a layer-by-layer mode due to thermodynamic 
reasons, the energy of the 3D islands being always lower than that 
of the complete  monolayers. The deposit atoms are almost in bulk-like 
positions. Equally important, the authors established that As greatly reduces 
the Si and Ge intermixing. This is another indication of the 
thermodynamic influence of the surfactant as the alloying also requires 
large atomic displacements.

\section{CONCLUSION}

The presence of a third element in the system unavoidably changes 
the kinetics of growth. Hence, we have to decide whether 
the kinetics is solely responsible for suppression of the 3D islanding, or 
the thermodynamics plays a decisive role, with the kinetics exerting an 
additional effect. We argue that this question could be answered by  
studying the temperature dependence of the capability of the surfactant to 
suppress 3D islanding. If the kinetics play a decisive role we should expect 
a change of the mode of growth from layer-by-layer at lower temperature to 3D 
(SK or VW) islands at higher temperature. The physical reason is that at high 
temperature the system is closer to equilibrium whereas at lower temperature 
the kinetic effects suppress the thermodynamic driving force for 3D islanding. 
This is most probably the case of the system Ge/Si(111) where one monolayer 
of the surfactant should be displaced by two monolayers of the overgrowth 
and the exchange process takes randomly place by single atoms. The S atoms 
do not press the overlayer atoms into bulk-like positions and formation of 
misfit dislocations or displacements of the edge atoms is not inhibited. We 
have seen that the growth is accompanied by typical kinetic phenomena as the 
kinetic regime of growth ($\chi  > 1$). On the contrary, the system 
Ge/Si(001) demonstrates the decisive role of the thermodynamics. The kinetics 
play an auxiliary role in the same direction. The physical reason for 
suppression of the 3D islanding is that a monolayer of the surfactant is 
displaced by a monolayer of the film during growth, and the exchange process 
occurs by dimers rather than by single atoms. As a result the overlayer 
atoms occupy nearly bulk-like positions provided by the outgoing surfactant 
dimers. Large atomic displacements are forcibly prohibited which in turn 
leads to complete wetting and a zero thermodynamic driving force for 3D 
islanding. It is thus clear that the third element in the system does not 
act as a surfactant at all.

We conclude that the relative weight of the effect of the third element on 
the thermodynamics and kinetics in semiconductor heteroepitaxy depends 
mostly on the mechanism of exchange of overlayer and surfactant atoms. 
Undoubtedly the third element changes the surface energies but it is the 
change of the relative adhesion of the overlayer to the substrate that is 
crucial from thermodynamic point of view. 

\acknowledgements

The author is greatly indebted to M. Paunov for the fruitfull discussion 
and the critical reading of the manuscript.

\begin{figure}
\caption{\label{chempot}
Schematic dependence of the film chemical potent- ial on the film thickness 
in number of monolayers for the three modes of growth denoted at each curve: 
VW - Volmer-Weber, SK - Stranski-Krastanov, and FM - Frank-van der Merwe.}\
\end{figure}

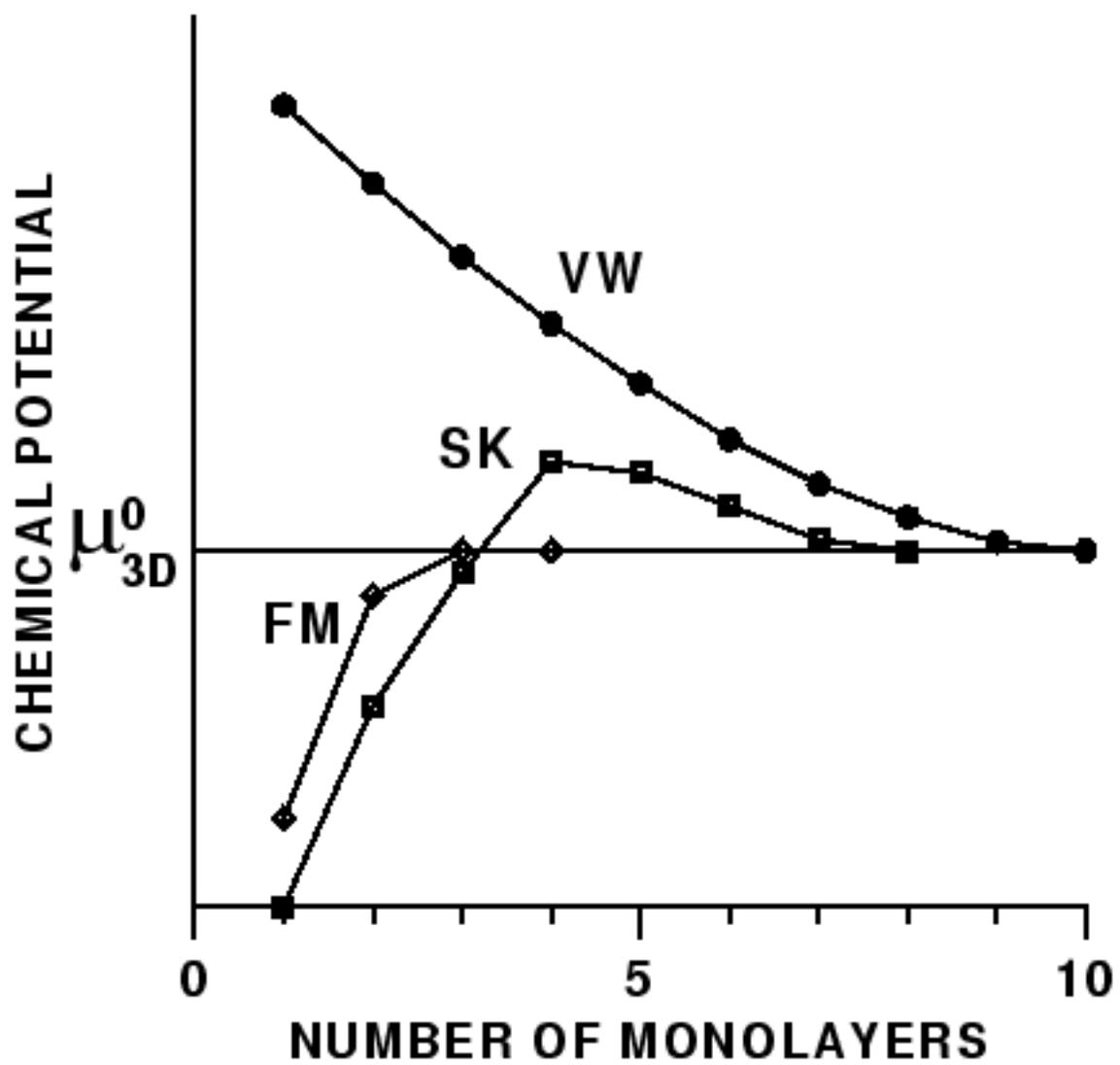
\end{document}